\documentclass[namedreferences]{solarphysics}
\usepackage{xcolor}
\usepackage{picture}
\usepackage[hyperref,optionalrh,solaromanenum]{spr-sola-addons} 
\usepackage{graphicx}                    
\usepackage{epstopdf}
\usepackage{amssymb,bm}                    
\usepackage{color}                       
\usepackage{breakurl}                         
\usepackage{hhline} 

\newcommand{\im}{\textrm{Im}}
\newcommand{\re}{\textrm{Re}}

\newcommand{\ii}{\textrm{i}}

\newcommand{\PB}{\Phi^{\tiny{\textrm{PB}}}} 
\newcommand{\EG}{\Phi^{\tiny{\textrm{LB}}}} 
\newcommand{\Phit}{K^{\tiny{\textrm{PB}}}} 
\newcommand{\Ht}{K^{\tiny{\textrm{LB}}}} 
\newcommand{\Htr}{\textrm{Re}[\Ht]}
\newcommand{\Phitr}{\textrm{Re}[\Phit]}
\newcommand{\Hti}{\textrm{Im}[\Ht]}
\newcommand{\Phiti}{\textrm{Im}[\Phit]}

\chardef\us=`\_

\usepackage{solaheader}	
\newcommand{\solareceiveddate}{21 July 2017} 
\newcommand{\solaaccepteddate}{5 January 2018}
\begin{document}

\begin{article}
\begin{opening}

\title{Ghost Images in Helioseismic Holography? Toy Models in a Uniform Medium}

\author[addressref=aff1,corref,email={yangd@mps.mpg.de}]{\inits{D.}\fnm{Dan}~\lnm{Yang}}
\address[id=aff1]{Max-Planck-Institut f\"ur Sonnensystemforschung, 37077 G\"ottingen, Germany}

\runningauthor{Dan Yang}
\runningtitle{Ghost Images in Helioseismic Holography?}

\begin{abstract}
Helioseismic holography is a powerful technique used to probe the solar interior based on estimations of the 3D wavefield. Porter--Bojarski holography, which is a well-established method used in acoustics to recover sources and scatterers in 3D, is also an estimation of the wavefield, and hence it has the potential to be applied to helioseismology. Here we present a proof of concept study, where we compare helioseismic holography and Porter--Bojarski holography under the assumption that the waves propagate in a homogeneous medium. We consider the problem of locating a point source of wave excitation inside a sphere. Under these assumptions, we find that the two imaging methods have the same capability of locating the source, with the exception that helioseismic holography suffers from ``ghost images'' (\textit{i.e.}, artificial peaks away from the source location). We conclude that Porter--Bojarski holography may improve the current method used in helioseismology. 
\end{abstract}

\end{opening}

\section{Introduction} 
Local helioseismology is a powerful tool used to probe the 3D interior of the Sun by exploiting the information  contained within the acoustic and surface-gravity waves observed at the  surface (see, \textit{e.g.}, \citealp{Gizon2005,Gizon2010}).  Helioseismic holography is one branch  of local helioseismology, which  aims at imaging the subsurface structure by estimating the wavefield inside the Sun \citep{Lindsey1997,Lindsey2000}. One significant achievement of  helioseismic holography has been the detection of active regions on the far-side of the Sun (far-side imaging: \citealp{Lindsey2000s}). The technique used in far-side imaging, known as phase-sensitive holography, has been validated  with synthetic data (see, \textit{e.g.}, \citealp{Hartlep2008,Birch2011,Braun2014}), and it is extensively used in studying active regions in the near hemisphere (\textit{e.g.}, \citealp{Braun2008,Braun2016}).

The fundamental concept of helioseismic holography is that the wavefield can be estimated by the so-called ``egression'', which is the back-propagation (in time) of the observed wavefield at the surface into the solar interior (see reviews by \citealp{Lindsey2000,Lindseybook}).  The egression can be understood in terms of Huygens' principle, whereby each point of a wavefront is considered a source, and the wavefield at a later time as a superposition of waves emitted from all of these point sources along the wavefront (see, \textit{e.g.}, \citealp{Landau_classical_field}). Specifically, each arbitrarily small  section of the observed wavefield can be regarded as a point source, and the egression as the sum of the back-propagated (in time) waves generated from all these point sources. Therefore, the egression  behaves in the same manner as the wavefield propagating backward in time. Furthermore, the propagation of the wavefield forward in time is known as the ``ingression''. The ingression can be understood by Huygens' principle in the same way as the egression, but for the waves that are forward-propagating in time.

Since Huygens' principle is frequently used in optics and acoustics, it is thus not surprising that techniques similar to the egression/ingression have been used in fields outside helioseismology. In ocean acoustics, \citet{Jackson1991} proposed an active method called phase conjugation, which used the same principle as the egression, to locate acoustic sources.  This method records the wavefield on an array of  detectors at a fixed surface, and then it creates a time-reversed (or a complex-conjugatation in the frequency domain) wave by treating the wavefield observed at each detector as a point source. \citet{Jackson1991} argued that the newly created wave will focus on the source of the original wave. This proposed method was later confirmed by experiments in the sea (see, \textit{e.g.}, \citealp{Kuperman1998,Song1998}).   

The rigorous  mathematical statement of Huygens' principle is the Helmholtz{--}Kirchhoff theorem, which states that the wavefield can be reconstructed in 3D space if both the wavefield and its normal derivative are recorded on an arbitrary closed surface (see, \textit{e.g.}, \citealp{Born1999}). However, the Helmholtz{--}Kirchhoff theorem is only valid for a source-free medium \citep{Porter1982}. This is not the case for the entirety of the solar interior, where the wavefield is stochastically and ubiquitously excited by near-surface convection.  In this case the medium is not source-free, and to our knowledge, no theory has been established thus far that can reconstruct the wavefield in 3D space directly from observations on a 2D surface.

Although the direct reconstruction of the wavefield is not possible,  acoustic sources and scatterers can be estimated  in 3D space for a medium that is not source-free. In acoustics, a well-established  technique known as Porter--Bojarski (PB) holography can  achieve this (see, \textit{e.g.}, \citealp{Porter1982, Devaney1985,Devaney2012}). 
PB holography is based on an integral that is slightly different from the Helmholtz--Kirchhoff theorem; however, the wavefield and its normal derivative are still required (see Section \ref{sec:kernel}).
Instead of reconstructing the wavefield, PB holography produces a 3D image (known as the PB hologram) from a closed  surface that is equivalent to the difference between the  wavefield and its time-reversal  \citep{Porter1969}.
Understanding the PB hologram can  be achieved through Huygens' principle. Specifically, the wavefield and its normal derivative recorded at a surface can be thought of as a dipole and a monopole source, respectively. 
This means that the PB hologram is also an estimation of the wavefield, and hence it has the potential to be applied to helioseismology in a similar manner to the egression.

With the possible application of PB holography, it is then natural to ask whether this new method better estimates the wavefield than the egression, and hence it improves the current imaging capabilities of heliseismic holography. Previous studies by  \citet{Skartlien2001,Skartlien2002} have shown that both the egression and the PB hologram are measurements of the local strength of acoustic sources and can be related to the sources via their respective sensitivity kernels. This allows us to refine the scope of the previous question; specifically, which method is more accurate at locating acoustic sources? Comparisons between these methods have been done in ocean acoustics, where \citet{Jackson1991} showed that both methods can locate the source. Hence, the authors concluded that the egression is the simplified version of the PB hologram, and chose to use the egression as their  preferred method since it is easier to implement. 
 In the case of helioseismology, however, the question of the optimal method  has yet to be answered. This is the goal of this article. 
Additionally, preliminary work by \citet{Lindsey2004} showed that helioseismic holography suffers from unintended mirror-like images (``ghost images'') due to the use of only a monopole source. An examination of ghost images in the egression and the PB hologram will also be a focus of this study.

In this article, we present a proof of concept study, where we compare the source-sensitivity kernels of helioseismic holography and PB holography by assuming waves are propagating in a homogeneous medium.  Specifically, we will examine which method is more accurate at locating acoustic sources, and hence estimating the wavefield. Additionally, we will examine the affect of observational coverage area on the kernels. This will provide an opportunity to improve the current method used in  helioseismology.  This article is organized as follows: Section \ref{sec:kernel} states the derivation of source-sensitivity kernels of helioseismic holography (egression) and  PB holography, Section \ref{sec:toy_model} states the toy model used in this paper, Section \ref{sec:results} compares the two methods with discussion and conclusions in Section \ref{sec:summary}.

\section{Source-Sensitivity Kernels}
\label{sec:kernel}
In this section, we  will present how  helioseismic holography and PB holography are related to acoustic sources. We  will work entirely in the temporal Fourier domain using the convention
\begin{equation}
f(\omega)=\int_{-\infty}^{+\infty} \mathrm{d} t F(t)  e^{\ii\omega t},
\end{equation} 
where $f(\omega)$  is the Fourier transform of a given function $F(t)$.

The egression $[\EG]$, as described by \citet{Lindsey1997} (LB), is one of the basic quantities used in helioseismic holography, 
\begin{equation}
\EG_A(\bm{r},\omega)= \int_A \mathrm{d}^2\bm{r}'G^*(\bm{r},\bm{r}', \omega)\Psi(\bm{r}', \omega),\label{eq:egression}
\end{equation}
where $\bm{r}$ denotes the focal point, $A$ is the coverage of the wavefield $\Psi(\bm{r}',\omega)$ at any point $\bm{r}'$ on the solar surface, and $G(\bm{r},\bm{r}',\omega)$ is the Green's function associated to a wave operator defined below with the asterisk denoting the complex conjugate. For simplicity, we will drop the  $\omega$ within the function's arguments for the remainder of this study. Further definitions and explanations concerning the Green's function will be given later in this section.

The PB hologram $\PB$ is defined by \citet{Devaney1985}, 
\begin{equation}
\PB_A(\bm{r})=\int_{A} \mathrm{d}^2\bm{r}'\left\{  \Psi(\bm{r}')\partial_{n'}\im G(\bm{r},\bm{r}')-\im G(\bm{r},\bm{r}') \partial_{n'}\Psi(\bm{r}')\right\}, \label{eq:pbhologram} 
\end{equation}
where $\partial_{n'}$ denotes the outward normal derivative with respect to $\bm{r}'$, and $\im G$ is the imaginary part of the Green's function.

In order to relate $\EG$ and $\PB$ to the acoustic sources,  we first need to determine the impulse response function [$G$, the Green's function] of the wave equation. Here we assume that the wavefield $[\Psi]$ is related to the sources through the application of a linear wave operator $[\mathcal{L}]$,
\begin{equation}
\mathcal{L}\Psi(\bm{r})=S(\bm{r}), \label{eq:eq_general}
\end{equation}
where $S(\bm{r})$ is the source function. For generality, we choose not to explicitly state $\mathcal{L}$ here. The Green's function is the impulse response of  Equation \ref{eq:eq_general}, and is defined as the solution to  
\begin{equation}
\mathcal{L}G(\bm{r},\bm{r}_s)=\delta(\bm{r}-\bm{r}_s),\label{eq:greens_general}
\end{equation}
where $\delta(\bm{r}-\bm{r}_s)$ is the Dirac delta function and $\bm{r}_s$ is the location of the source. One property of the Green's function, which is crucial to deriving source sensitivity kernels, is that it can be used to solve Equation \ref{eq:eq_general} through
\begin{equation}
\Psi(\bm{r'})=\int_{\mathbb{R}^3} \mathrm{d}^3\bm{r}_s G(\bm{r}',\bm{r}_s)S(\bm{r}_s). \label{eq:greens_technique}
\end{equation}

Through expansion of  $\Psi$ in Equation \ref{eq:egression} with the definition in Equation \ref{eq:greens_technique}, the egression becomes
\begin{equation}
\EG_A(\bm{r})=\int_A \mathrm{d}^2\bm{r}' G^*(\bm{r},\bm{r}') \int_{\mathbb{R}^3} \mathrm{d}^3\bm{r}_s G(\bm{r}',\bm{r}_s)S(\bm{r}_s),
\end{equation}
and through a change in the order of integration, one obtains the definition for the source sensitivity kernel $[\Ht]$ for the egression
\begin{equation}
\EG_A(\bm{r})=\frac{1}{4\pi}\int_{\mathbb{R}^3} \mathrm{d}^3\bm{r}_s \Ht_A(\bm{r},\bm{r}_s) S(\bm{r}_s), \label{eq:egression_source}
\end{equation}
where
\begin{equation}
\Ht_A(\bm{r},\bm{r}_s)=4\pi\int_A \mathrm{d}^2\bm{r}' G^*(\bm{r},\bm{r}')  G(\bm{r}',\bm{r}_s).\label{eq:egression_kernel}
\end{equation}

We note that the $4\pi$ factor is included here such that $\Ht$ possesses a desired near-unitary amplitude for this study.

The same procedure is repeated for the derivation of the PB hologram by  expanding $\Psi$ in Equation \ref{eq:pbhologram} with Equation \ref{eq:greens_technique} and  changing the order of integration;
\begin{eqnarray}
\PB_A(\bm{r})&=&\frac{1}{4\pi}\int_{\mathbb{R}^3} \mathrm{d}^3\bm{r}_s \Phit_A(\bm{r},\bm{r}_s)S(\bm{r}_s),\label{eq:pb_source}\\
\Phit_A(\bm{r},\bm{r}_s)&=&4\pi\int_{A} \mathrm{d}^2\bm{r}' \left\{  G(\bm{r}',\bm{r}_s)\partial_{n'}\im G(\bm{r},\bm{r}') -\im G(\bm{r},\bm{r}') \partial_{n'} G(\bm{r}',\bm{r}_s)\right\}, \label{eq:pb_kernel} \nonumber\\
\end{eqnarray}
where $\Phit_A$ is the source sensitivity kernel of the PB hologram.

A comparison of  the egression and the PB hologram requires only the knowledge of their respective source-sensitivity kernels, whereas details of the source function are not needed.  Therefore, we will examine the source-sensitivity kernels of the two imaging methods in this study. We will compare the two source kernels under simplifying assumptions about the medium in which the waves propagate. We note that, in practice, the egression power $|\EG|^2$ is used to estimate the location of acoustic sources, since the wavefield in the Sun is stochastically excited (see, \textit{e.g.}, \citealp{Lindsey1997,Hanson2015}). Therefore, we will also compare the squared modulus of the source kernels.

\section{Toy Model: Waves in a Homogeneous Medium}  
\label{sec:toy_model}

With Equations \ref{eq:egression_kernel} and \ref{eq:pb_kernel} in hand, we require the computation of the Green's functions in order to determine the source-sensitivity kernels. In general, a Green's function can be obtained numerically for any given linear operator $[\mathcal{L}]$. However, as stated in the introduction, we examine a homogeneous medium and as such the  Green's function can be computed analytically.

We consider this homogeneous medium in $\mathbb{R}^3$ space with a constant sound speed $c=10^5$ m$\cdot\textrm{s}^{-1}$, and we adopt the linear wave operator,
\begin{equation} 
\mathcal{L}=k^2+\nabla^2,\ k= \frac{\omega+\textrm{i}\gamma}{c} \label{eq:helm},
\end{equation}
where $k$ is the wavenumber, $\omega$ is the angular frequency and $\gamma$ is the damping rate. The solution of Equation \ref{eq:greens_general} with the above wave operator and free boundary condition is given by 
\begin{equation}
G(\bm{r},\bm{r}')=-\frac{1}{4\pi}\frac{\exp\left(\ii k|\bm{r}-\bm{r}'|\right)}{|\bm{r}-\bm{r}'|},\label{eq:4}
\end{equation}
which is also known as the \textit{outgoing free-space} Green's function \citep{Born1999}. In this study, we set the damping rate $[\gamma]$ to be $0.1\,\%$ of the angular frequency, and we use a Cartesian coordinate system $(x,y,z)$ with its origin at the center of a sphere $\textrm{V}_{\odot}$ with the radius  $\textrm{R}_\odot=696\ \textrm{Mm}$. 

Current observational capabilities mean that  we can only observe the wavefield  on  a fraction of the solar surface. To study the consequence of  this limitation on observations, we will examine both the case where the entire surface is observed and the case where only a fraction of the solar surface is observed. In these cases, we assume the sources are located along the $z$-axis, the coverage is symmetric with respect to the $z$-axis and is centered above the North Pole $(0,0,\textrm{R}_\odot)$. Under these assumptions, $\Ht$ and $\Phit$ are  axisymmetric about the $z$-axis.

\section{Results}
\label{sec:results}
\subsection{Source-Sensitivity  Kernels at 3~mHz}
\label{sec:3mHz}
\begin{figure}[!ht]
\centering
\includegraphics[width= \textwidth]{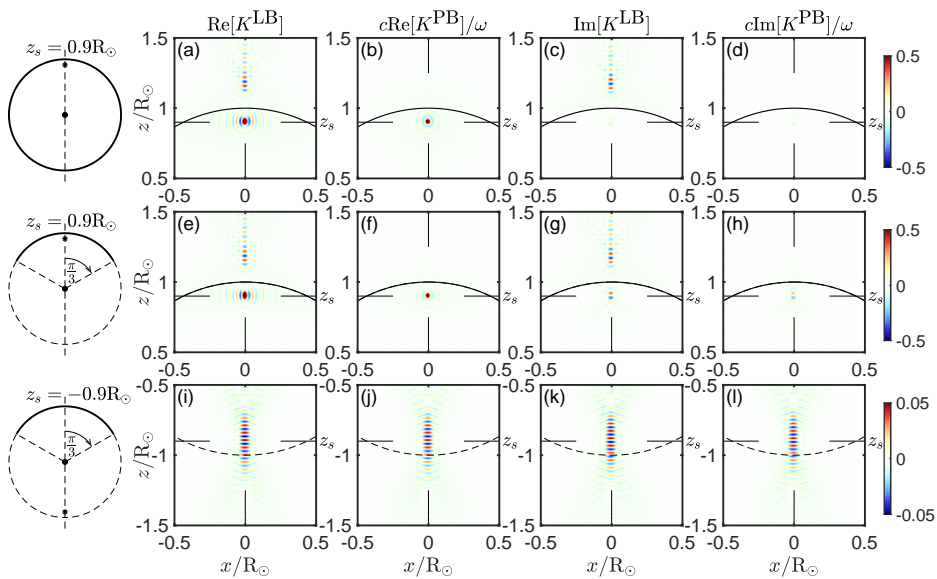}
 \caption{2D slices of the real and imaginary parts of  $\Ht$ and $\Phit$  through the $z$-axis, when the entire surface (top row) and 60 degrees around the North Pole (middle and bottom rows) can be  observed. The location of the source  is at the focal point of the cross hairs (solid black lines) in each plot, being $z_s=0.9\textrm{R}_\odot$ in the top and middle rows and  $z_s=-0.9\textrm{R}_\odot$ in the bottom row. A simple  geometry plot is given on the left of each row, with the source location (asterisk), the solar center (big dot), solar surface (dashed lines), and coverage area (solid arc) on top of it.}\label{fig:map_summary}
\end{figure}

\begin{figure}[!ht]
\begin{center}
\includegraphics[width= \textwidth, trim= {0 0 0  {-0.14\textwidth}}]{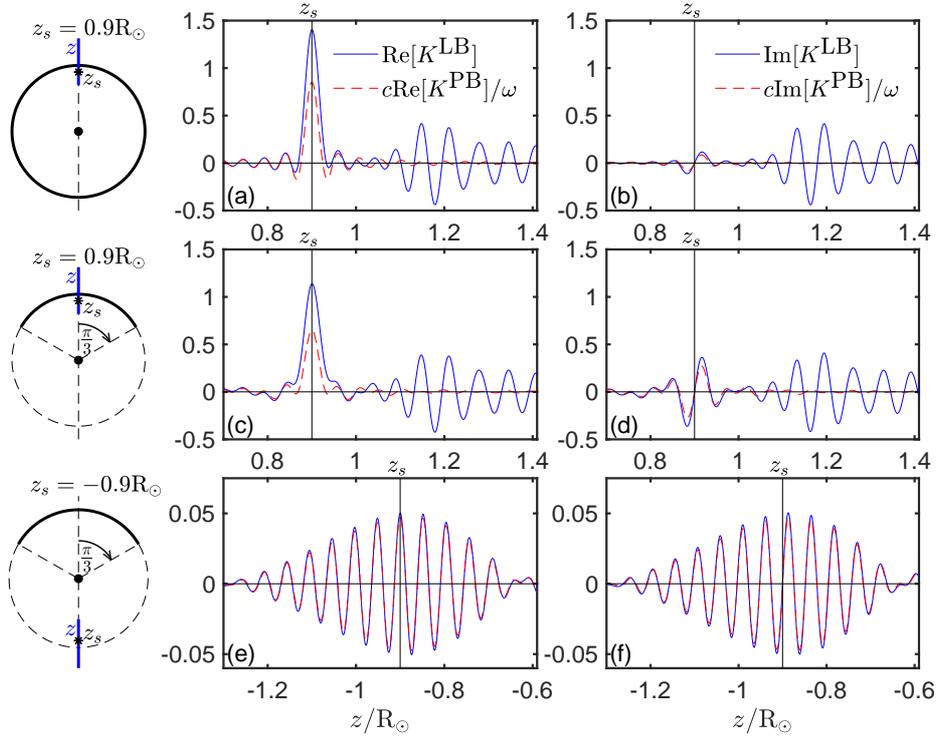}
 \caption{1D slices of $\Ht$ (blue solid) and $\Phit$ (red dashed)  shown in Figure \ref{fig:map_summary} along the $z$ axis. The real parts are plotted in panels a, c, and e, and  the imaginary parts in panels b, d, and f. The geometry is shown on the left with the addition of the slice length (blue line).}\label{fig:slice_1d_radial}
\end{center}
\end{figure} 

Here we examine the source-sensitivity kernels [$\Ht$ and $\Phit$] at a frequency of $\omega/2\pi=3$~mHz. Figure \ref{fig:map_summary} shows 2D slices of both the real and imaginary parts of  $\Ht$ and  $c\Phit/\omega$ through the $z$-axis, for the source locations at $z_s=0.9\textrm{R}_\odot$ (panel a to h) and $-0.9\textrm{R}_\odot$ (panel i to l) on the $z$ axis.  The factor $c/\omega$ is added to  $\Phit$ so  that the kernel is dimensionless like the egression. The first row of panels show the source-sensitivity kernels under the assumption that the entire surface is observed. The remaining rows show the kernels assuming a coverage of 60~degrees  from the North Pole. Considering all panels,  both $\Htr$ and $\Phitr$ peak at the source, while in comparison $\Hti$ and $\Phiti$ are negligible if the source is located at the near side. These results demonstrate that both $\Htr$ and $\Phitr$ can locate the source in both of these coverage geometries. In the case of far-side located sources, all of the kernels have became less localized. While these kernels can locate the sources, we also see that the egression kernels (both the real and imaginary parts) have ``ghost images'' above the surface,  while the PB holograms do not. These ghost images appear as peaks at points away from the source location. We note that in this work we also observed ghost images below the surface  in the egression, when the source is above the surface.   This suggests that the egression cannot distinguish sources from below and above the surface, since one can not differentiate sources and ghosts. The PB holograms do not suffer from this problem. Further explanations and discussions concerning the ghost images will be given in Section \ref{sec:discussion_ghost}.

For a more focused comparison of the kernels in Figure \ref{fig:map_summary}, Figure \ref{fig:slice_1d_radial} shows 1D slices of $\Ht$ and $\Phit$ along the $z$-axis with  the real parts shown in panel a, c, and e, and the imaginary parts in panel b, d, and f.  $\Htr$ and $\Phitr$ again have peaks at the source location. Here we see that despite the coverage geometry, $\Hti$ and $\Phiti$ are always zero at the source location with the peaks seen in Figure \ref{fig:map_summary} surrounding the source location. This suggests that  $\Hti$ and $\Phiti$ cannot pinpoint the exact source location. In the case of far-side located sources, $\Ht$ and  $\Phit$ are both highly oscillatory and non-localized, and hence recovering the source location may be problematic with observations at a single frequency. \\

\begin{figure}[!ht]
\begin{center}
\includegraphics[width=0.6 \textwidth, trim= {0 0 0  {-0.14\textwidth}}]{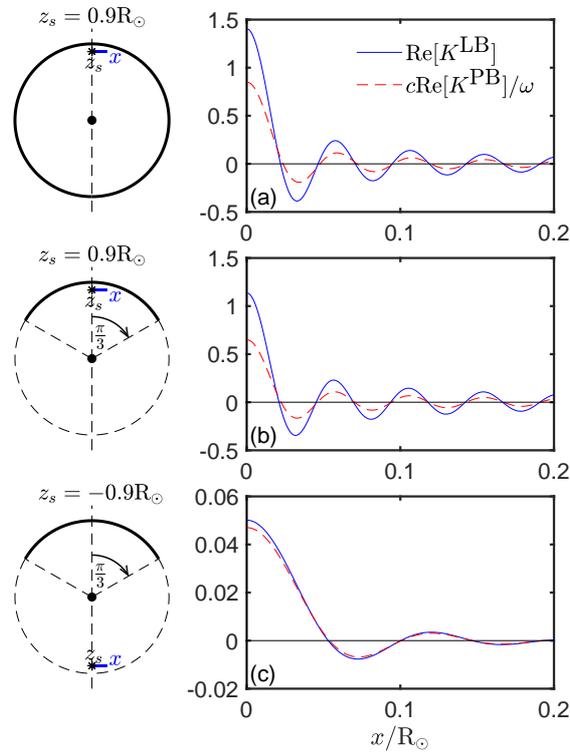}
 \caption{1D slices of $\Ht$ (blue solid) and $\Phit$ (red dashed)  in Figure \ref{fig:map_summary} along the line  perpendicular to the $z$-axis. The imaginary parts are negligible compared to the real parts, and they are not shown in the figure. Due to axial symmetry, only half of the slice is shown. The geometry of the slices is shown on the left with the slice length shown in blue.}\label{fig:slice_1d_nonradial}
\end{center}
\end{figure}
Figure \ref{fig:slice_1d_nonradial} shows 1D slices of $\Ht$ and $\Phit$  along a line that is perpendicular to the $z$-axis and passes through the source. Due to axial symmetry, only half of the slice is plotted. Here the imaginary parts of $\Ht$ and $\Phit$ are not shown since they are negligible compared to the real parts. Unlike the vertical slices, both $\Ht$ and $\Phit$ do not have ghost images, and they are less oscillatory when the source is located at the far-side.

\subsection{Kernels Averaged over Frequency}
\label{sec:averaged_kernels}
Specifically, for observations at a single frequency  both methods are highly oscillatory and non-localized for the far-side located source, and the egression suffers from ghost images for the near-side located source.  One possible solution to the above issues is to average kernels over a number of  frequencies, since the ghost images for the near-side located source and the side-lobes for the far-side located source may peak at different locations for different frequencies.

\begin{figure}[!ht]
\begin{center}
\includegraphics[width=\textwidth, trim= {0 0 0  {-0.14\textwidth}}]{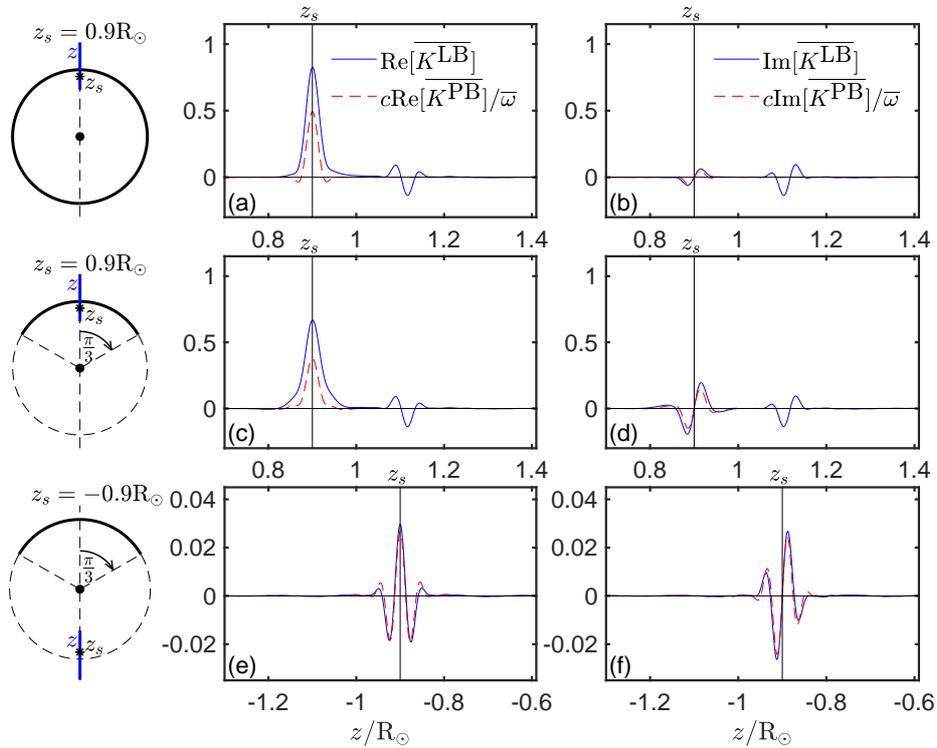}
\caption{1D slices of kernels averaged over frequencies using a Gaussian weight function centered at 3 mHz with a standard deviation of 1 mHz. 
 Here we use a bar to denote the averaged quantity.
The observational coverage $A$ is depicted in the left panels (thick arcs). The frequency averaging reduces the amplitudes of the ghost images for the near-sided source and the side-lobes for the source on the far-side.} \label{fig:averaged_kernels_radial_bw4mhz}
\end{center}
\end{figure}
Figure \ref{fig:averaged_kernels_radial_bw4mhz} shows 1D vertical slices of $\Ht$ and $\Phit$ averaged from 41 frequencies equally distributed from 1 to 5 mHz. A Gaussian weight function centered at 3~mHz and with a standard deviation of $1 $~mHz had been applied for the averaging. From these results, it is clear that averaging kernels over frequency  reduces the amplitude of the ghost images for the near-sided source and the side-lobes for the far-side located source. The  averaged kernels along the horizontal direction are similar to the kernels with  a single frequency, and as such they are not shown here.

\begin{figure}[!ht]
\begin{center}
\includegraphics[width=\textwidth, trim= {0 0 0  {-0.14\textwidth}}]{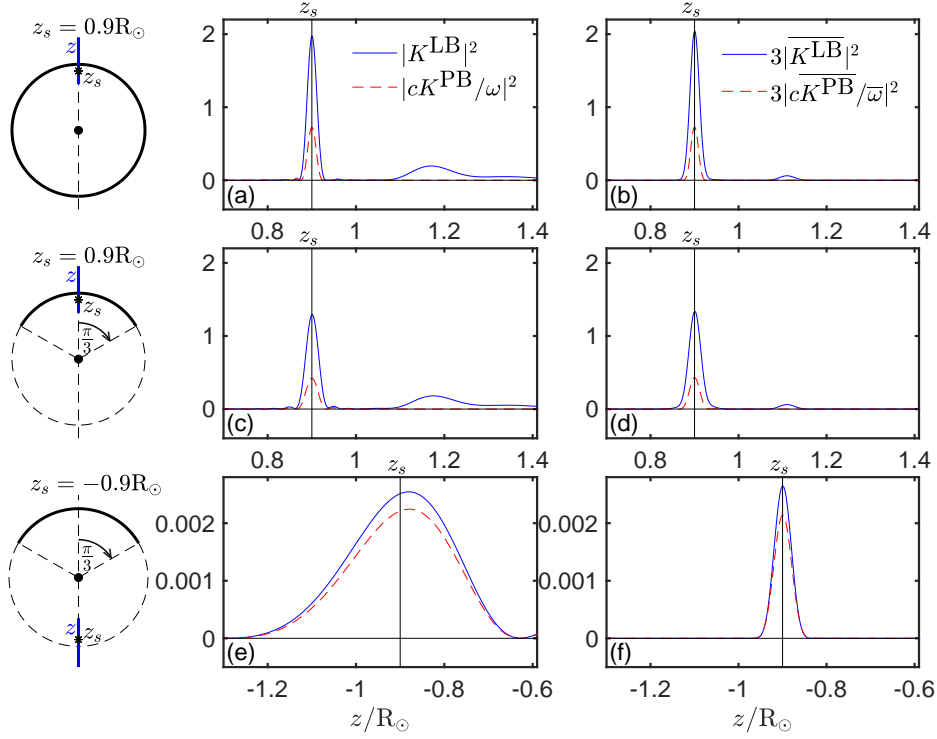}
\caption {1D slices of $|\Ht|^2$ and $|\Phit|^2$ in a plane containing the $z$-axis. The left panels are for a single frequency of 3~mHz. The right panels are for averages over frequencies using a Gaussian weight function centered at 3 mHz with a standard deviation of 1 mHz.   We see that averaging the kernels over frequencies reduces the amplitude of the ghosts when the source is located on the near-side, and it improves the spatial resolution when the source is located on the far-side.}
\label{fig:averaged_kernels_power_radial_bw4mhz}
\end{center}
\end{figure}

As mentioned in Section \ref{sec:kernel}, the  egression power, which is related to the source covariance via $|\Ht|^2$, has been used in observations as estimations of the acoustic sources. Therefore, the effect of averaging $|\Ht|^2$ and $|\Phit|^2$ over different frequencies is also of great interest. Figure \ref{fig:averaged_kernels_power_radial_bw4mhz} shows a comparison of $|\Ht|^2$ and $|\Phit|^2$  with or without averaging over different frequencies. Only the slices along the vertical direction are shown, as the difference between $|\Ht|^2$ ($|\Phit|^2$) from a single frequency at 3~mHz and averaged from 1 to 5~mHz along the horizontal direction is small. In the vertical direction,  averaging $|\Ht|^2$ and $|\Phit|^2$ over different frequencies reduces the amplitude of the ghosts when the source is located on the near-side, and improves the spatial resolution when the source is located on the far-side.

\subsection{Dependence of the Spatial Resolution on the Coverage}
\label{sec:spatial_resolution}
The results thus far have shown that both of the methods can locate the source, though the egression has the complication of ghost peaks. The question then arises of how well do these methods resolve the sources with differing observational coverages?  
We define the spatial resolution the egression and the PB hologram as the full width at the half maximum (FWHM) of $|\Ht|^2$ and $|\Phit|^2$ respectively.  From Figures  \ref{fig:slice_1d_radial} and  \ref{fig:slice_1d_nonradial}  we can see that both of the methods behave differently  in the vertical and horizontal directions, and as such the FWHM in these two directions are considered separately. 
Additionally, to quantify the affect of averaging kernels over different frequencies, both the FWHM for kernels from a single frequency at $3$~mHz and the value averaged over frequencies from 1 to 5~mHz are considered.          

\begin{figure}[!ht]
\begin{center}
\includegraphics[width= \textwidth, trim= {0 0 0  {-0.14\textwidth}}]{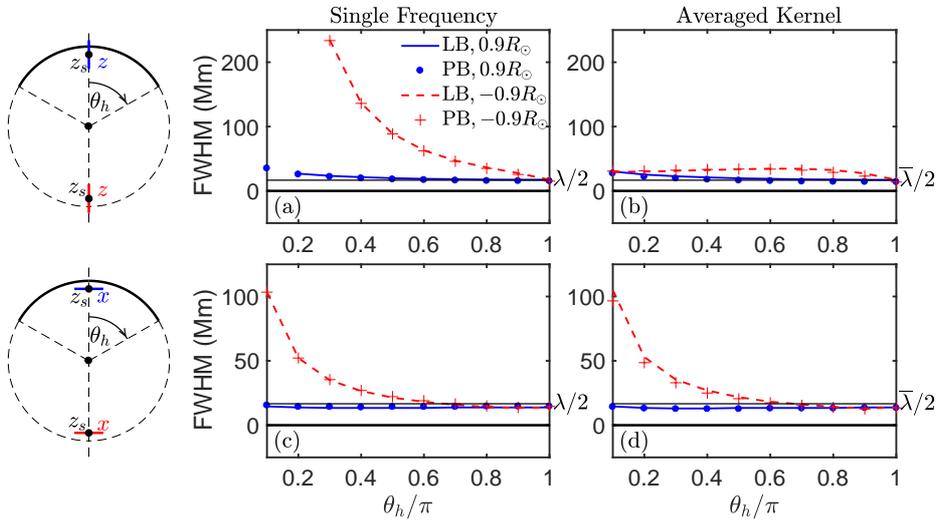}
 \caption{The full width at half maximum (FWHM)  of $|\Ht|^2$ and $|\Phit|^2$  as a function of the angle $[\theta_h]$, which defines the observational coverage (cap of area $A$). The FWHM along two directions, horizontal  and vertical,  are shown in the case of a near-side located source ($z_s=0.9\textrm{R}_\odot$, blue) and a far-side located source ($z_s=-0.9\textrm{R}_\odot$, red). The theoretical resolution limit of $\lambda/2$ is also shown with a horizontal black line.  Additionally, both the results for the kernels at a single frequency $\omega/2\pi = 3$~mHz and the frequency-averaged kernels are shown on the left and right columns, respectively. }\label{fig:Resolution}
\end{center}
\end{figure}
Figure \ref{fig:Resolution} shows the FWHM of $|\Ht|^2$ and $|\Phit|^2$ as a function of the  angle $[\theta_h]$, which defines the observational coverage (cap of area $[A]$). The FWHM along vertical (top row) and horizontal (bottom row) directions are considered in the case of kernels at 3~mHz (left column) and averaged from 1 to 5~mHz (averaged kernel, right column). At 3~mHz, the difference between the two imaging methods is small, which implies that either method has the same capability to resolve the source. Furthermore, the FWHM is close to the resolution limit despite the size of the coverage area when the source is located at the near-side ($z_s=0.9\textrm{R}_\odot$). When the source is located at the far-side ($z_s=-0.9\textrm{R}_\odot$), the resolution improves (FWHM decreases) with increasing coverage. When averaging over different frequencies, a clear improvement of the spatial resolution can be found along the vertical direction when the source is located on the far-side, while the spatial resolution is almost the same as before for other cases.

\section{Discussion}
\label{sec:summary}
\subsection{Ghost Images in the Egression}
\label{sec:discussion_ghost}

\begin{figure}[!ht]
\begin{center}
\includegraphics[width=\textwidth, trim= {0 0 0  {-0.22\textwidth}}]{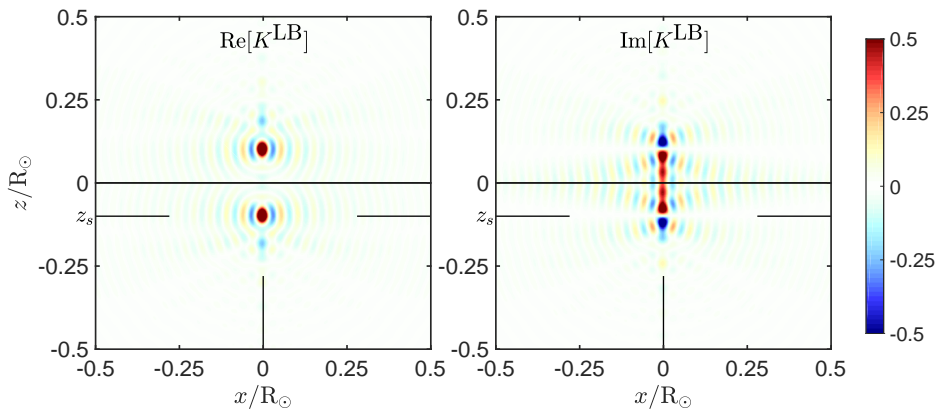}
\caption{2D slice of the real (left panel) and  imaginary (right panel) part of  $\Ht$ through the $z$-axis when the wavefield is observed at the $z=0$ plane. Here the coverage is a circle that centered at the origin and with the radius of $\textrm{R}_\odot$.  The source is located along the $z$-axis at $z_s = -0.1\textrm{R}_\odot$ and is indicated by the focus of  the cross hairs in each plot.} \label{fig:ghost_plane}
\end{center}
\end{figure} 
The appearance of ghost images in the egression can be understood by Huygens' principle, whereby each arbitrarily small section of the observed wavefield is regarded as a point source, and the egression as a superposition of the back-propagated (in time) waves generated from all the point sources. Furthermore, each newly created wave is spherically  symmetric with respect to its source location in a homogeneous medium, and thus it will propagate in all directions with the same behavior. This is the cause of the ghost images. A clear example of this is when the wavefield is recorded on a plane, where all of the newly created waves are symmetric  with respect to the recording surface, and hence the egression  will focus on both the source location and its counterpart on the other side of the surface (see Figure \ref{fig:ghost_plane} for an example).  When the wavefield is observed on a sphere, however, the newly created waves are no longer symmetric with respect to the surface, and the ghost images show a complicated diffraction  pattern due to the interference among the newly created waves (Figure \ref{fig:map_summary}). This provides a simple explanation for the ghost images seen in the egression above the surface (see also \citealp{Lindsey2004}).

The PB hologram does not suffer from ghost images like the egression, since it includes not only a monopole source but also a dipole source, which is not symmetric with respect to the source location. Additionally, the amplitudes of the monopole and dipole sources are chosen  such that the PB hologram only focuses on the source location.

Future work should include a solar-like density stratification to confirm this simple explanation. The sharp drop in density at the solar surface leads to a reflection of the waves below 5.3 mHz, which is not captured in our toy model.

We also note that \citet{Lindsey2005a,Lindsey2005b} proposed that ghost images may explain the presence of phase anomalies observed around active regions in phase-sensitive holography. For further implications and  discussions about the ghost images in helioseismic holography, we refer readers to \citet{Lindsey2004}.

\subsection{The Ingression and the PB Hologram}
In this study, we have not considered the ingression in our analysis. The ingression is an equally important quantity used in helioseismic holography, which is an estimation of where the wavefield converges to by propagating the wavefield forward in time \citep{Lindsey1997}.
So far, we have considered the wavefield $\Psi$ as diverging away from the source. However, to compare the ingression and the PB hologram, a wavefield that converges from infinity to the source location is desired. Such a wavefield can be achieved by considering the wavefield diverging from the source as before, but with the reversed sign in time, \textit{i.e.} $\Psi(\bm{r},-t)$. 
In the frequency domain, this time-reversal corresponds to taking the complex conjugate. Additionally, the wave number [$k$] in the wave equation is also conjugated and thus the wavefield decays when propagating towards the source location \citep{Devaney2012}. In this case, the ingression is
\begin{equation}
\EG_{A,-}(\bm{r},\omega)=\int_{A} \mathrm{d}^2\bm{r}'G(\bm{r},\bm{r}',\omega)\Psi^*(\bm{r}',\omega),
\end{equation} 
and the PB hologram becomes 
\begin{equation}
\PB_{A,-}(\bm{r},\omega)=\int_{A} \mathrm{d}^2\bm{r}' \{  \Psi^*(\bm{r}',\omega) \partial_{n'}\im G(\bm{r},\bm{r}',\omega)-\im G(\bm{r},\bm{r}',\omega) \partial_{n'}\Psi^*(\bm{r}',\omega)\}.\label{eq:pbhologram_minus} 
\end{equation}
 We can see that $\EG_-$ and  $\PB_- $ are simply the complex conjugates of $\EG$ and $\PB$. Since we have discussed the real and imaginary parts and the power of $\PB$ and $\EG$ separately in the results, those of $\EG$ ($\PB$) will be the same as $\EG_-$ ($\PB_-$). In particular, $\EG_-$ will also have ghost images while $\PB_-$ will not, and $|\EG_-|^2$ and $|\PB_-|^2$ will have the same spatial resolution when imaging acoustic sources.

\subsection{Application to Stereoscopic Helioseismology}
\label{sec:Stereoscopic}
\begin{figure}[!ht]
\begin{center}
\includegraphics[width=\textwidth, trim= {0 0 0  {-0.16\textwidth}}]{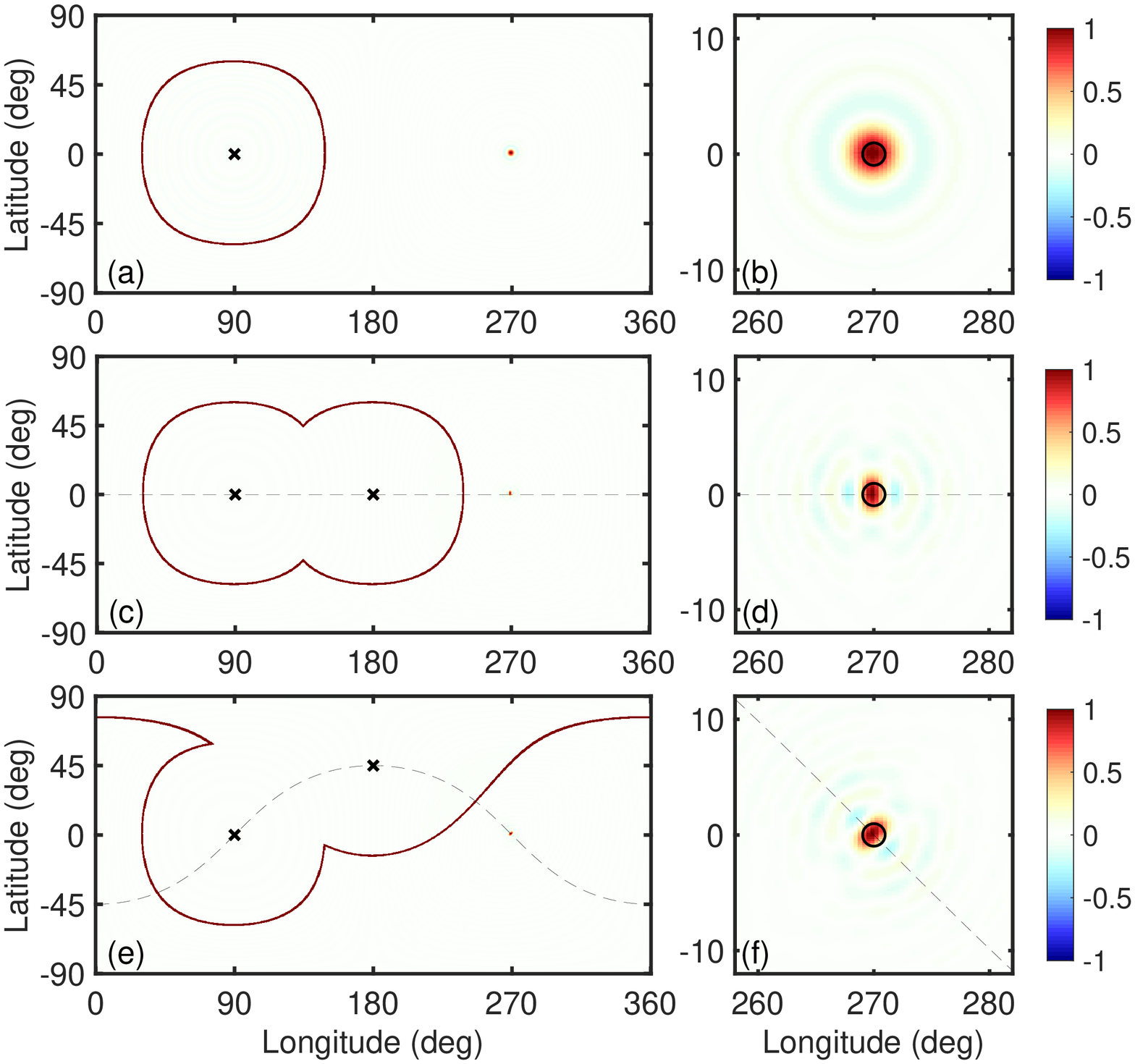}
\caption{2D slices of $\re[\Phit]$ at the solar surface for observations from a single spacecraft (top row), two spacecraft in the Ecliptic (middle row), and two spacecraft with one in the Ecliptic and the  other at  $45^\circ$ inclination (bottom row). Here the source is located 0.7~Mm below the surface at $270^\circ$ longitude  along the Equator, and the plots are shown after divided by the maximum value of $\re[\Phit]$. We note that $\im[\Phit]$ is negligible, and as such is not shown here. We plot $\re[\Phit]$ at the entire surface on the left column, where the boundary of the coverage is marked by a red curve and the point below the spacecraft by a cross. Additionally, a zoom of images around the source location are shown on the right column, where a circle centered above the source location and with a diameter of the wavelength is added on each plot.
We can see a clear improvement of the spatial resolution when a second spacecraft is added. Furthermore, a preferred direction that possesses  higher spatial resolution  is found along the great circle (black-dashed line)  that goes through the points below the two spacecraft.} \label{fig:solar_sterescopy}
\end{center}
\end{figure}

Results in Section \ref{sec:spatial_resolution} showed that the spatial resolution of the hologram on the far-side  increases as the coverage area increases.
Thus the resolution has a fundamental limit when observing from a single vantage point.  It has been suggested to combine observations from two or several vantage points to increase the observation coverage and therefore to improve the spatial resolution (and signal-to-noise ratio) of holography. Stereoscopic helioseismology  is believed to be our best chance to probe the subsurface structure in the polar regions and the deep convection zone, which is crucial for  understanding  the 11-year solar cycle (see, \textit{e.g.}, \citealp{Ruzmaikin2003}). 
This conjecture, however, has not been studied in detail.

Figure~\ref{fig:solar_sterescopy} shows the PB hologram at the solar surface with a Dirac delta source located 0.7~Mm below the surface at $270^\circ$ longitude along the Equator. Different coverage geometries are considered in the case of  a single spacecraft (top row), two spacecraft in the ecliptic (middle row), and two spacecraft with one in the Ecliptic and the other at  $45^\circ$ inclination (bottom row). Here only the real parts of the PB hologram are shown, since the imaginary parts are negligible.  The results  show a clear improvement of the spatial resolution when a second spacecraft is added, whereas the FWHM along the Equator is about two times smaller than that of a single spacecraft. Additionally, the spatial resolution is increased along the great circle (black-dashed line)  at the intersection of the plane that contains the center of the sphere and the two spacecraft.

Stereoscopic helioseismology might be implemented in future space missions such as \textit{Solar Orbiter} and \textit{Solar Activity Far Side Investigation} (see, \textit{e.g.}, \citealp{Sekii2015} and references therein) together with observations collected from the ground (\textit{Global Oscillation Network Group}) or from near-Earth orbit (\textit{Solar Dynamics Observatory}). In particular,
 the \textit{Polarimetric and Helioseismic Imager} onboard \textit{Solar Orbiter} is to be launched soon and will provide high-resolution line-of-sight velocity and continuum intensity at the photosphere, which are suitable for  helioseismic studies (\citealp{Woch2007, Muller2013,Loeptien2015}). 
The  orbit of \textit{Solar Orbiter} will have a period of 168 days during the nominal mission and reach a heliographic latitude of up to  $25^{\circ}$  ($35^{\circ}$ during an extended mission) \citep{Muller2013}. 
This means that \textit{Solar Orbiter} will cover a large range of spacecraft--Sun--Earth angles to test stereoscopic helioseismology.

\section{Outlook}
In this article we found that helioseismic holography and PB holography are similar techniques, with the exception that the egression and the ingression suffer from ghost images.  In principle, we could apply the PB holograms to  phase-sensitive holography by replacing  the  egression and the ingression with the appropriate $\PB$ and $\PB_-$. Our toy model suggests that the PB holograms will improve current helioseismic holography since they do not suffer from ghost images. However additional modeling work is needed. Future studies should consider random acoustic sources and  scatterers.  Furthermore, the computations must be carried out in a solar-like stratified background medium.  Finally, in order to implement PB holography  a method for determining the normal derivative of the wavefield needs to be developed.

%
\begin{acks}
I thank Laurent Gizon, Aaron C. Birch, Damien Fourier, and Chris S. Hanson for their useful insights and comments on the article. I also thank the referee, Charles Lindsey, for his thoughtful comments. This work was supported by the International Max Planck Research School (IMPRS) for Solar System Science at the University of G\"ottingen.
\end{acks}


\bibliographystyle{spr-mp-sola}
\bibliography{holo.bib}  

\end{article} 
\end{document}